\documentstyle[epsfig]{article}

\begin{document}

\title{IMPROVED CONSTRAINTS ON WIMPS FROM THE INTERNATIONAL GERMANIUM EXPERIMENT IGEX}

\author{}

\date{}

\maketitle

\begin{center}

A. Morales$^{a}$\footnote{Corresponding author:
amorales@posta.unizar.es}, C.E. Aalseth$^{b}$, F.T. Avignone
III$^{b}$, R.L. Brodzinski$^{c}$, S. Cebri\'{a}n$^{a}$, E.
Garc\'{\i}a$^{a}$, \\
I.G. Irastorza$^{a}$\footnote{Present address: CERN, EP Division,
CH-1211 Geneva 23, Switzerland}, I.V. Kirpichnikov$^{d}$, A.A.
Klimenko$^{e}$, H.S. Miley$^{c}$, \\ J. Morales$^{a}$, A. Ortiz de
Sol\'{o}rzano$^{a}$, S.B. Osetrov$^{e}$, V.S. Pogosov$^{f}$, J.
Puimed\'{o}n$^{a}$, J.H.\ Reeves$^{c}$, \\ M.L. Sarsa$^{a}$,
A.A. Smolnikov$^{e}$, A.G. Tamanyan$^{f}$, A.A. Vasenko$^{e}$,
S.I. Vasiliev$^{e}$, J.A. Villar$^{a}$
\end{center}

\begin{center}
\begin{em}

$^{a}$Laboratory of Nuclear and High Energy Physics, University of
Zaragoza, 50009 Zaragoza, Spain
\\
$^{b}$University of South Carolina, Columbia, South Carolina 29208
USA
\\
$^{c}$Pacific Northwest National Laboratory, Richland, Washington
99352 USA
\\
$^{d}$Institute for Theoretical and Experimental Physics, 117 259
Moscow, Russia
\\
$^{e}$Institute for Nuclear Research, Baksan Neutrino Observatory,
361 609 Neutrino, Russia
\\

$^{f}$Yerevan Physical Institute, 375 036 Yerevan, Armenia \\

\end{em}
\end{center}


\abstract{One IGEX $^{76}$Ge double-beta decay detector is
currently operating in the Canfranc Underground Laboratory in a
search for dark matter WIMPs, through the Ge nuclear recoil
produced by the WIMP elastic scattering. A new exclusion plot,
$\sigma$(m), has been derived for WIMP-nucleon spin-independent
interactions. To obtain this result, 40 days of data from the IGEX
detector (energy threshold $E_{thr} \sim 4$ keV), recently
collected, have been analyzed. These data improve the exclusion
limits derived from all the other ionization germanium detectors
in the mass region from 20~GeV to 200~GeV, where a WIMP supposedly
responsible for the annual modulation effect reported by the DAMA
experiment would be located. The new IGEX exclusion contour
enters, by the first time, the DAMA region by using only raw data,
with no background discrimination, and excludes its upper left
part. It is also shown that with a moderate improvement of the
detector performances, the DAMA region could be fully explored.}

\section{Introduction}

Experimental observations and robust theoretical arguments have
established that our universe is essentially non-visible, the
luminous matter scarcely accounting for one per cent of the
critical density of a flat universe ($\Omega=1$). The current
prejudice is that the universe consists of unknown especies of
Dark Energy ($\Omega_{\Lambda}\sim 70\%$) and Dark Matter
($\Omega_{M}\sim 25-30\%$) of which less than $\sim 5\%$ is of
baryonic origin. Most of that Dark Matter is supposed to be made
of non-baryonic particles filling the galactic halos, at least
partially according to a variety of models. Weak Interacting
Massive (and neutral) Particles (WIMPs) are favourite candidates
to such non-baryonic components. The lightest stable particles of
supersymmetric theories, like the neutralino, describe a
particular class of WIMPs.

WIMPs can be detected by measuring the nuclear recoil produced by
their elastic scattering off target nuclei in a suitable detector
\cite{Mor99}. In particular, non-relativistic ($\sim 300$ km/s)
and heavy ($10 - 10^3$ GeV) galactic halo WIMPs could make a Ge
nucleus recoil with a few keV, at a rate which depends on the type
of WIMP and interaction. Only about 1/4 of this energy is visible
in the detector. Because of the low interaction rate (which ranges
from 10 to $10^{-5}$ counts/kg/day according to the SUSY model and
the choice of parameters) and the small energy deposition (from a
few to $\sim$100 keV), the direct search for particle dark matter
through scattering off nuclear targets requires ultralow
background detectors with very low energy thresholds.



Germanium detectors have reached one of the lowest background
levels of any type of detector and  have a reasonable ionization
yield (nuclear recoil ionization efficiency relative to that of
electrons of the same kinetic energy) ranging from 20\% to 30\%
depending on the nuclear recoil energy . Thus, with sufficiently
low energy thresholds, they are attractive devices for WIMP
searches. That is the case for IGEX.

This paper presents new WIMPs constraints in the cross-section
WIMP-nucleon versus WIMP mass plot, derived from a germanium
detector (enriched up to 86 \% in $^{76}$Ge) of the IGEX
collaboration, which improve previous limits obtained with Ge
ionization detectors, and enter by the first time the so-called
DAMA region (corresponding to a WIMP supposedly responsible for
the annual modulation effect found in the DAMA experiment
\cite{Ber99}) without using mechanisms of background rejection,
but relying only in the ultra-low background achieved.



\section{Experiment}
The IGEX experiment \cite{Aal,Gon99}, optimized for detecting
$^{76}$Ge double-beta decay, has been described in detail
elsewhere. One of the IGEX detectors of 2.2 kg, enriched up to 86
\% in $^{76}$Ge, is being used to look for WIMPs interacting
coherently with the germanium nuclei.
Its active mass is $\sim2.0$~kg, measured with a collimated source
of $^{152}$Eu.
The full-width at half-maximum (FWHM) energy resolution is
2.37~keV at the 1333~keV line of $^{60}$Co.
Energy calibration and resolution measurements were made
periodically using the lines of $^{22}$Na and $^{60}$Co.
Calibration for the low energy region was extrapolated using the
X-ray lines of Pb. The uncertainty induced by this extrapolation
in the determination of the energy values in the threshold region
has been estimated to be smaller than 0.1 keV, as deduced from the
check of linearity performed systematically --with this and other
detectors of IGEX-- along several years since the arrival of the
detector to the underground facility, when the activation peaks
(at about 10 keV) were still visible.

The Ge detector and its cryostat were fabricated following
state-of-the-art ultralow background techniques and using only
selected radiopure material components (see Ref \cite{Aal,Mor00}).
The first-stage field-effect transistor (FET) of the detector is
mounted on a Teflon block a few centimeters from the central
contact of the germanium crystal. The protective cover of the FET
and the glass shell of the feedback resistor have been removed to
reduce radioactive background. This first-stage assembly is
mounted behind a 2.5-cm-thick cylinder of archaeological lead to
further reduce the background. Further stages of preamplification
are located at the back of the cryostat cross arm, approximately
70 cm from the crystal. The IGEX detectors have preamplifiers
modified for the pulse-shape analysis used in the double-beta
decay searches.

The detector shielding has been recently modified with respect to
that of the previous set-up of Ref. \cite{Mor00}, improving the
external neutron shielding and increasing the thickness of lead
surrounding the detector. The shielding is now as follows: the
innermost shield consists of about 2.5 tons of 2000-year-old
archaeological lead of ancient roman origin (having $<9$~mBq/kg of
$^{210}$Pb($^{210}$Bi), $< 0.2$~mBq/kg of $^{238}$U, and
$<0.3$~mBq/kg of $^{232}$Th) forming a cubic block of 60~cm side.
The detector is fitted into a precision-machined chamber made in
this central core, which minimizes the empty space around the
detector available to radon. Nitrogen gas, at a rate of
140~l/hour, evaporating from liquid nitrogen, is forced into the
small space left in the detector chamber to create a positive
pressure and further minimize radon intrusion. The archaeological
lead block is surrounded, at its turn, by 20 cms of lead bricks
made from 70-year-old low-activity lead ($\sim 10$ tons) having
$\sim 30$~Bq/kg of $^{210}$Pb. The whole lead shielding forms a
1-m cube, the detector being surrounded by not less than 40-45~cm
of lead (25 cm of which is archaeological). A 2-mm-thick cadmium
sheet surrounds the main lead shield, and two layers of plastic
seal this central assembly against radon intrusion. A cosmic muon
veto covers the top and sides of the shield, except where the
detector Dewar is located. The veto consists of BICRON BC-408
plastic scintillators 5.08 cm $\times$ 50.8 cm $\times$ 101.6 cm
with surfaces finished by diamond mill to optimize internal
reflection. BC-800 (UVT) light guides on the ends taper to 5.08 cm
in diameter over a length of 50.8 cm and are coupled to Hamamatsu
R329 photomultiplier tubes. The anticoincidence veto signal is
obtained from the logical OR of all photomultiplier tube
discriminator outputs with a count rate lesser than 40 Hz (i. e.
using a threshold which allows to include events with a poor light
collection). An external neutron moderator 40~cm thick formed by
polyethylene bricks and borated water tanks completes the shield.
The entire shield is supported by an iron structure resting on
noise-isolation blocks. The experiment has an overburden of 2450
m.w.e., which reduces the muon flux to a (measured) value of $2
\times 10^{-7} \rm cm^{-2} \rm s^{-1}$.

\begin{figure}[ht]
\centerline{ \epsfxsize=14cm \epsffile{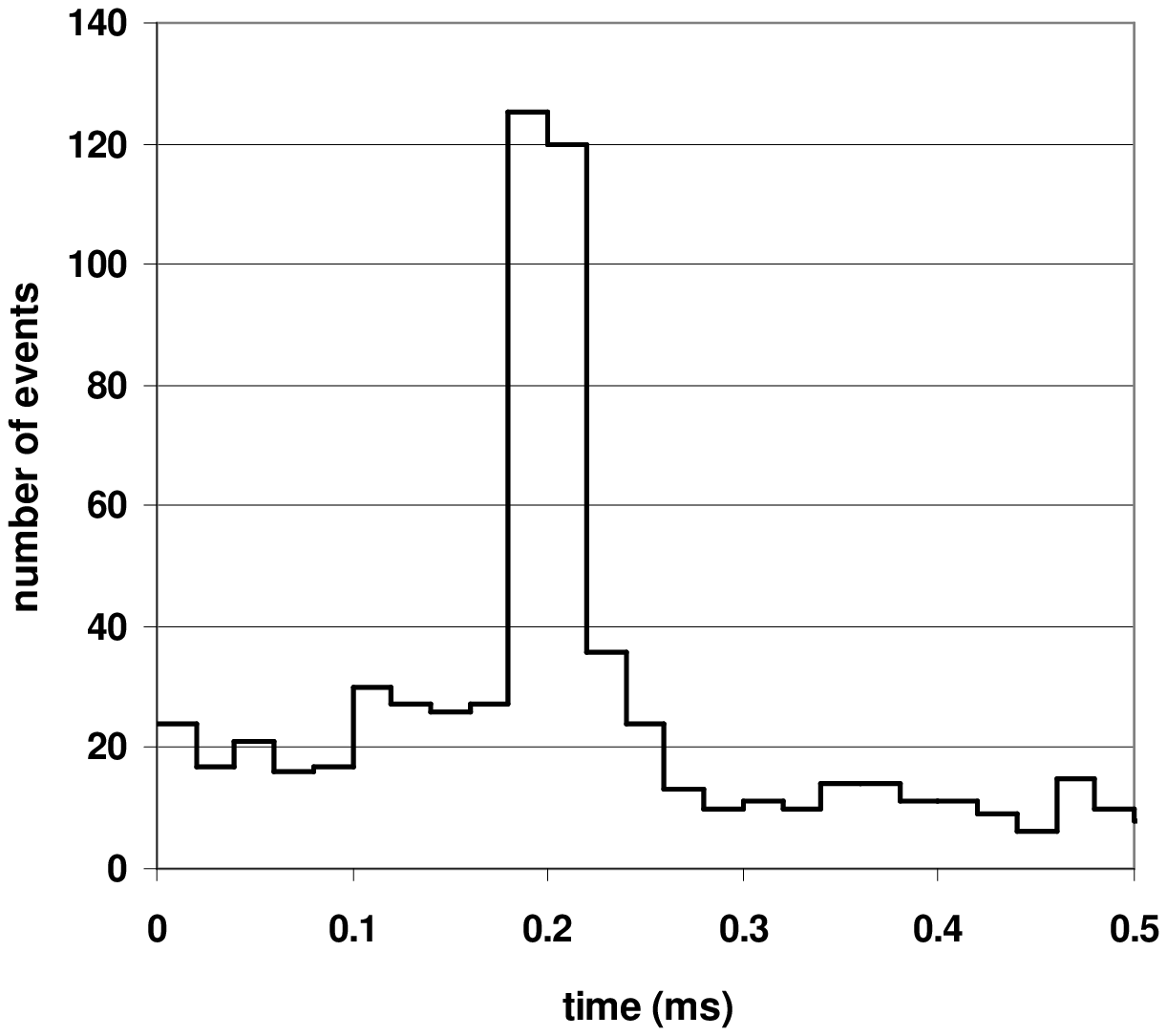} }
 \caption{Distribution of the time after last veto event.
 The distribution is centered at $\sim 200 \mu$s as expected
 (see text).} \label{veto_window}
\end{figure}
\begin{figure}[ht]
\centerline{ \epsfxsize=14cm \epsffile{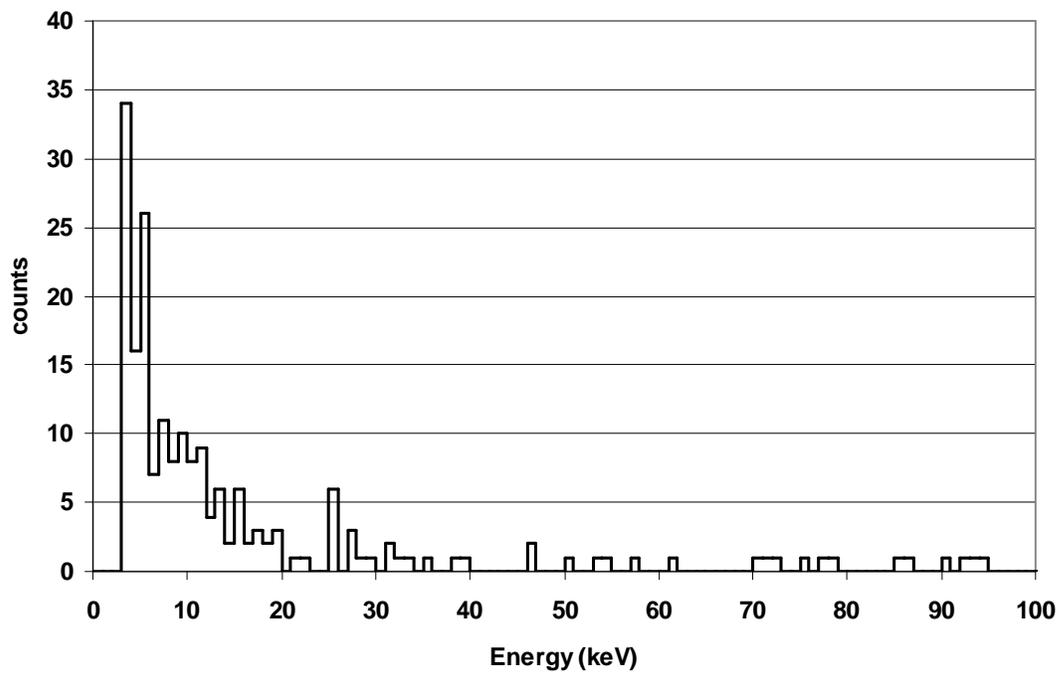} }
 \caption{Distribution at low energies of the events
 rejected by the veto system.} \label{veto_spectrum}
\end{figure}

The data acquisition system for the low-energy region used in this
WIMP search is based on standard NIM electronics. It has been
implemented by splitting the normal preamplifier output pulses of
the detector and routing them through two Canberra 2020 amplifiers
having different shaping times enabling noise rejection as first
applied in Ref \cite{Jmor}. The minimum settled ratio of the two
amplitudes processed with different shaping time depends on the
energy and in the present case it ranges from 0.8 (at 4 keV) to
0.99 (at 50 keV). These amplifier outputs are converted using 200
MHz Wilkinson-type Canberra analog-to-digital converters,
controlled by a PC through parallel interfaces. For each event,
the arrival time (with an accuracy of 100~$\mu$s), the elapsed
time since the last veto event (with an accuracy of 20~$\mu$s),
and the energy from each ADC are recorded.
Figure \ref{veto_window} shows the time-after-last-veto
distribution of the events. Notice that it reflects properly the
200 $\mu$s delay included in the main trigger of the acquisition
system for the computer to decide whether it acquires or not the
digitized pulse after knowing its energy. The muon veto
anticoincidence was done off-line with a software window up to
240$\mu$s.
The probability of rejecting non-coincident events is less than
0.01. The rejected veto-coincident events amount up to about the
5\% of the total rate and are distributed in the low energy region
as shown by the Figure \ref{veto_spectrum}.

In addition, the pulse shapes of each event before and after
amplification are recorded by two 800 MHz LeCroy 9362 digital
scopes. These are analyzed one by one by means of a method based
on wavelet techniques which allows us to assess the probability of
this pulse to have been produced by a random fluctuation of the
baseline. The method requires the calculation of the wavelet
transform of $f(x)$, the recorded pulse shape after amplification:

\begin{equation}\label{def_wavelet_trans}
  [W_\psi f](a,b) = \frac{1}{\sqrt{a}}\int_{-\infty}^\infty f(x)
  \psi\left(\frac{x-b}{a}\right)dx
\end{equation}

where the "mexican hat" wavelet function $\psi(x) \propto
(1-x^2)\exp\left(-\frac12x^2\right)$ was chosen for our purposes.
Following expression (\ref{def_wavelet_trans}), a two-parameter
function $[W_\psi f](a,b)$ was numerically obtained for each
event. The relative maxima of this function were calculated, the
highest one corresponding to the event pulse and the others to
random fluctuations of the baseline. It was proven that the
distribution of the values of the wavelet transform at these
points follows an exponential. By comparing the maximum
corresponding to the event pulse with this exponential one can
calculate the probability $P$ for the first maximum to belong to
the distribution of the other maxima. This value is the final
output of the analysis for each event and is interpreted as the
probability of the main pulse of being randomly generated by the
fluctuations of the baseline.

\begin{figure}[ht]
\centerline{ \epsfxsize=14cm \epsffile{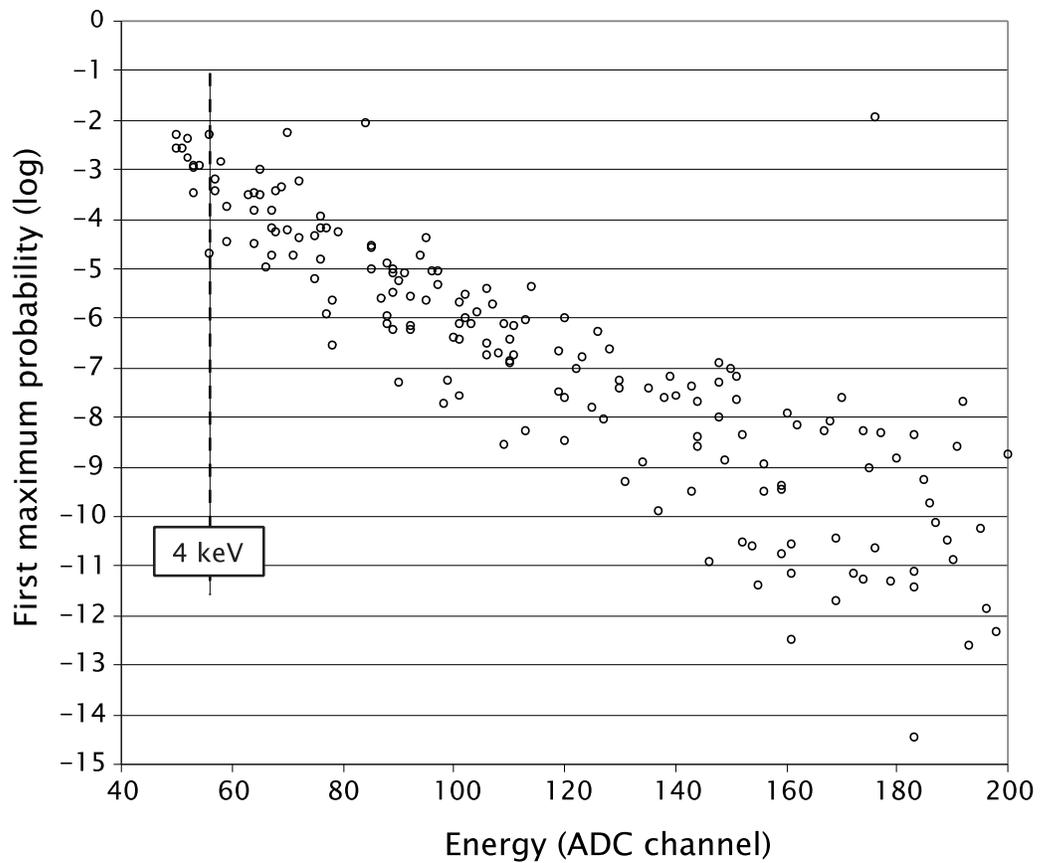} }
 \caption{Scatter plot of the probability assigned to each
 event by the wavelet technique (described in the text) versus
 energy for a calibration set of data. } \label{wavelet_cal}
\end{figure}

\begin{figure}[ht]
\centerline{ \epsfxsize=14cm \epsffile{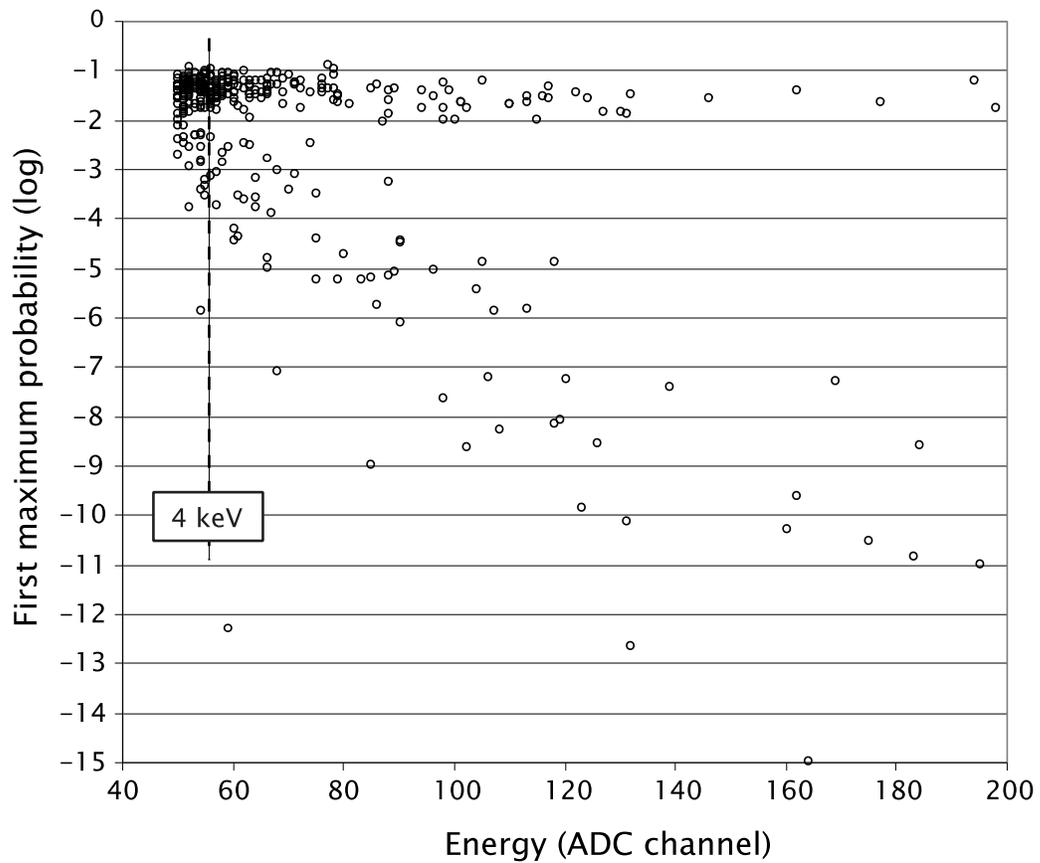} }
 \caption{Same as Figure \protect\ref{wavelet_cal} but for
 background data. The populations of noise and data are well
 separated above 4 keV.} \label{wavelet_bkg}
\end{figure}

In order to fix the rejection criterion this method was applied to
a calibration set of data. The result is shown in Figure
\ref{wavelet_cal} where the probability obtained with this method
versus energy is presented for each event. The same plot but for a
background set of data is shown in Figure \ref{wavelet_bkg}. From
these plots a criterion of $P<0.01$ can be defined to distinguish
the two populations of noise and data. Although it is hard to
quantify the loss of efficiency with the available statistics, the
figures show that it is very small for events above 4 keV.

It is worth mentioning that in spite of its good efficiency, this
technique was not able, by itself alone, to improve the previous
low energy background presented in Ref. \cite{Mor00}. Therefore,
we concluded that noise and microphonics does not contribute
substantially to such background and, consequently, the reduction
of background presented in the next section is attributed to the
changes in the shielding.

\section{Results and prospects}

\begin{figure}[ht]
\centerline{ \epsfxsize=14cm \epsffile{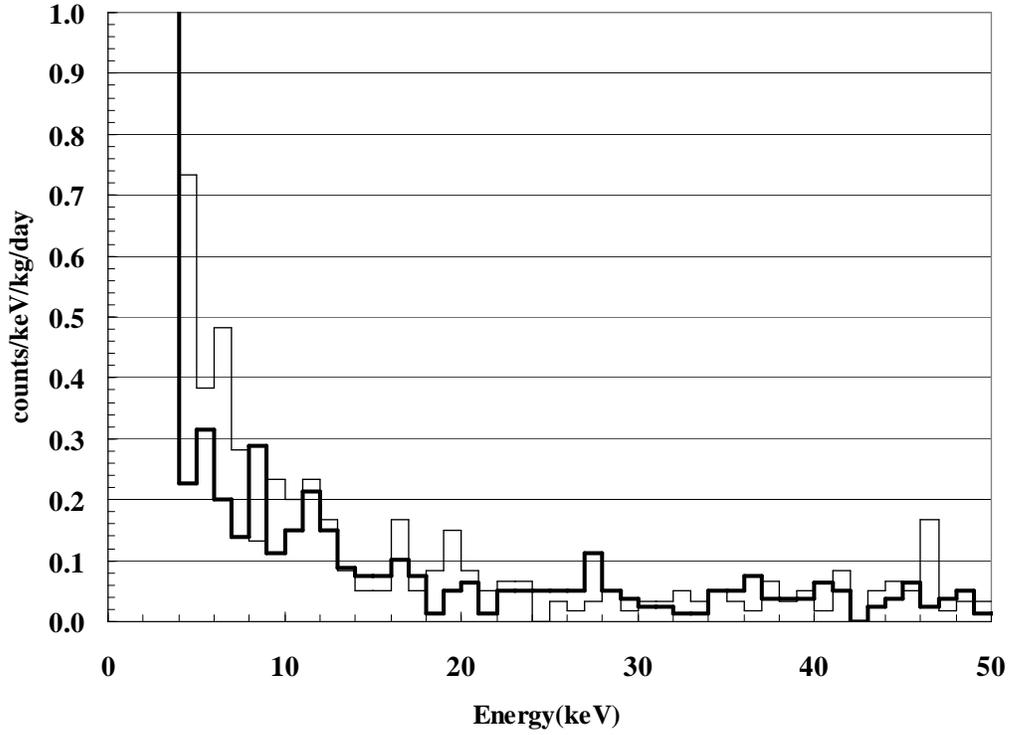} }
 \caption{Normalized low energy spectrum of the IGEX RG-II detector
 corresponding to the 80~kg~d presented in this paper (thick line)
 compared to the previous 60~kg~d spectrum published in \cite{Mor00}.}
 \label{dm-ig-1}
\end{figure}

\begin{figure}[ht]
\centerline{ \epsfxsize=14cm \epsffile{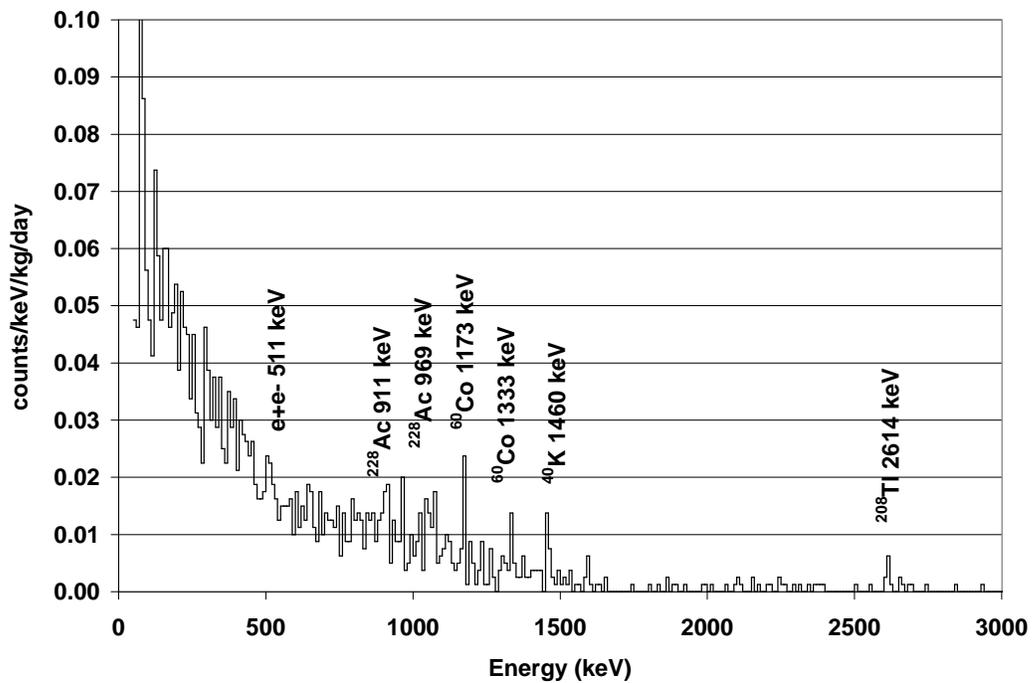} }
 \caption{High energy region of the spectrum of the IGEX RG-II detector
 corresponding to the 80~kg~d presented in this paper.} \label{spcHE}
\end{figure}

The results presented in this paper are from a recent run with the
modified shielding and analysis system previously described. They
correspond to an exposure of Mt=80 kg~days. The spectrum obtained
is shown in Figure~\ref{dm-ig-1} compared with the previous IGEX
published spectrum of Ref. \cite{Mor00}. The numerical data are
also given in Table~\ref{tab-ig-1}. The high energy region up to 3
MeV is shown in Figure \ref{spcHE}.

The energy threshold of the detector is 4 keV and the FWHM energy
resolution at the 75 keV Pb X-ray line was of 800 eV. The
background rate recorded was $\sim 0.21$ c/keV/kg/day between
4--10~keV, $\sim 0.10$ c/keV/kg/day between 10--20~keV, and $\sim
0.04$ c/keV/kg/day between 25--40~keV. As it can be seen, the
background below 10 keV has been substantially reduced (about a
factor 50\%) with respect to that obtained in the previous set-up
\cite{Mor00}, essentially due to the improved shielding (both in
lead and in polyethylene-water).
As was stressed before, this reduction was not due to the
implementation of the Pulse Shape Analysis, which suggests that
the neutrons could be an important component of the low energy
background in IGEX.

The exclusion plots are derived from the recorded spectrum in
one-keV bins from 4~keV to 50~keV, by requiring the predicted
signal in an energy bin to be less than or equal to the (90\%
C.L.) upper limit of the (Poisson) recorded counts. The derivation
of the interaction rate signal supposes that the WIMPs form an
isotropic, isothermal, non-rotating halo of density $\rho =
0.3$~GeV/cm$^{3}$, have a Maxwellian velocity distribution with
$\rm v_{\rm rms}=270$~km/s (with an upper cut corresponding to an
escape velocity of 650~km/s), and have a relative Earth-halo
velocity of $\rm v_{\rm r}=230$~km/s. The cross sections are
normalized to the nucleon, assuming a dominant scalar interaction.
The Helm parameterization \cite{Eng91} is used for the scalar
nucleon form factor. To compare the IGEX exclusion plots with that
derived from the Heidelberg-Moscow data \cite{Bau}, the recoil
energy dependent ionization yield used is the same that in Ref.
\cite{Bau}, E$_{\rm vis}=0.14\ (\rm E_{recoil})^{1.19}$.

\begin{figure}[ht]
\centerline{ \epsfxsize=14cm \epsffile{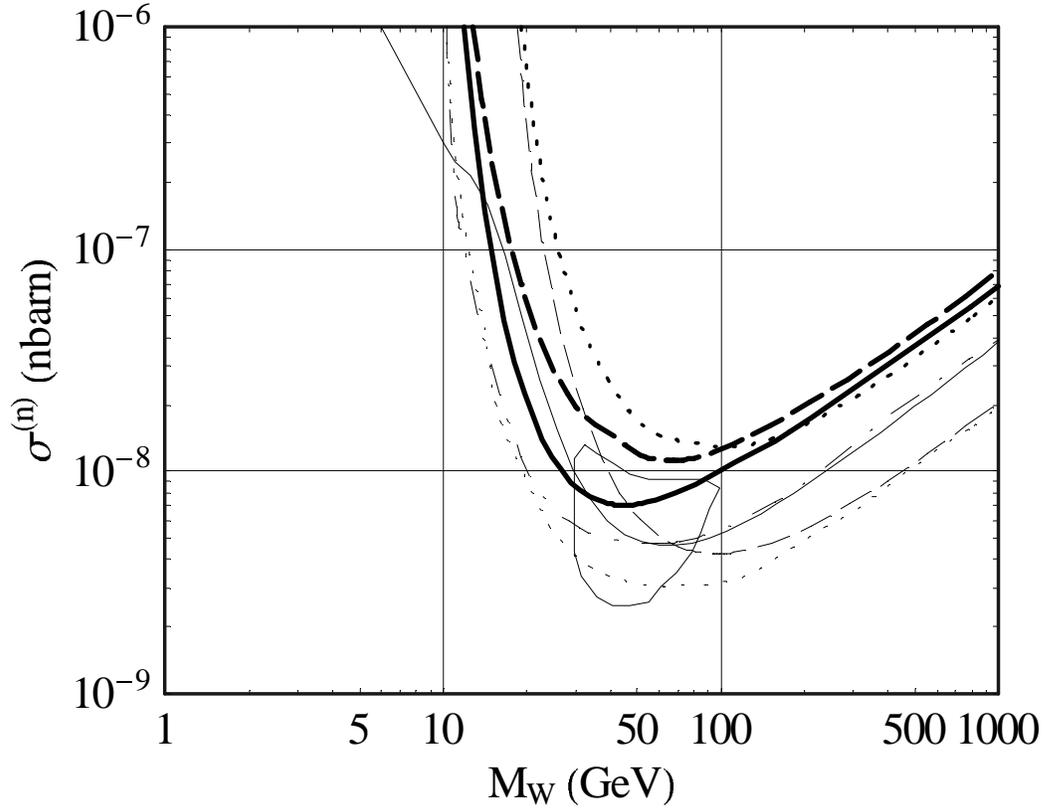} }
 \caption{IGEX-DM exclusion plot for spin-independent
interaction obtained in this work (thick solid line) compared with
the previous exclusion obtained by IGEX-DM \protect\cite{Mor00}
(dashed thick line) and the last result obtained by the
Heidelberg-Moscow germanium experiment \cite{Bau} (dotted line)
recalculated from the original spectrum with the same set of
hypothesis and parameters. The closed line corresponds to the
(3$\sigma$) annual modulation effect reported by the DAMA
collaboration (including NaI-1,2,3,4 runnings) \cite{Ber99}. The
thin solid line is the exclusion line obtained by DAMA NaI-0
\cite{Ber96} by using Pulse Shape Discrimination. The two other
experiments which have entered the DAMA region are also shown:
EDELWEISS \cite{Benoit:2001} (thin dashed line) and the CDMS
exclusion contour (thin dotted line) \cite{Abusaidi:2000} and its
expected sensitivity \cite{Sadoulet:2001} (thin dot-dashed line).
} \label{dm-ig-2}
\end{figure}

The exclusion plot derived in this way is shown in
Fig.~\ref{dm-ig-2} (thick solid line). It improves the IGEX-DM
previous result (thick dashed line) as well as that of the other
previous germanium ionization experiments (including the last
result of Heidelberg-Moscow experiment \cite{Bau} --thick dotted
line--) for a mass range from 20~GeV to 200~GeV, which encompass
that of the DAMA mass region.
In particular, this new IGEX result excludes WIMP-nucleon
cross-sections above 7 $\times 10^{-9}$ nb for masses of $\sim$50
GeV and enters the so-called DAMA region \cite{Ber99} where the
DAMA experiment assigns a WIMP candidate to their found annual
modulation signal. IGEX excludes the upper left part of this
region. That is the first time that a direct search experiment
without background discrimination, but with very low (raw)
background, enters such region.
Also shown for comparison are the contour lines of the other
experiments, CDMS \cite{Abusaidi:2000} and EDELWEISS
\cite{Benoit:2001} (thin dashed line), which have entered that
region, as well as the DAMA region (closed line) corresponding to
the $3\sigma$ annual modulation effect reported by that experiment
\cite{Ber99} and the exclusion plot obtained by DAMA NaI-0 (thin
solid line) \cite{Ber96} by using statistical pulse shape
discrimination. A remark is in order: for CDMS two contour lines
have been depicted according to a recent recommendation
\cite{Sadoulet:2001}, the exclusion plot published in Ref.
\cite{Abusaidi:2000} (thin dotted line) and the CDMS expected
sensitivity contour \cite{Sadoulet:2001} (thin dot-dashed line).



\begin{figure}[ht]
\centerline{ \epsfxsize=14cm \epsffile{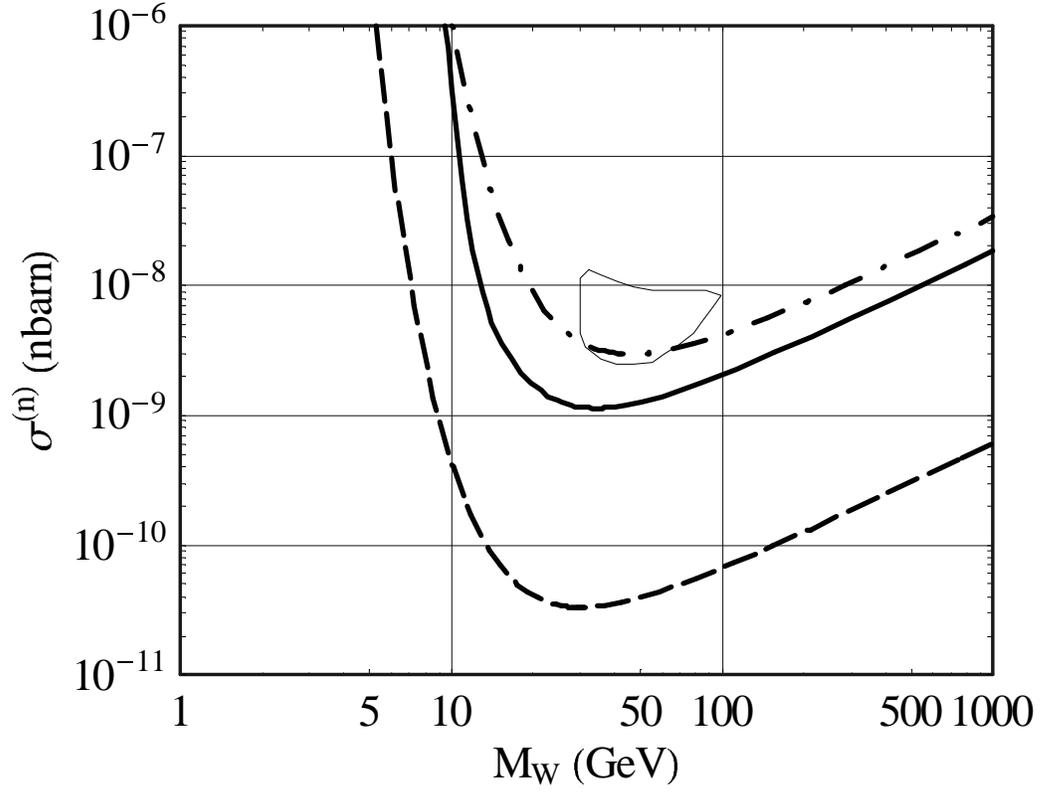} }
 \caption{IGEX-DM projections are shown for
 a flat background rate of 0.1~c/keV/kg/day (dot-dashed line) and 0.04~c/keV/kg/day (solid line) down
 to the threshold at 4 keV, for 1~kg~year of exposure.
 The exclusion contour expected for GEDEON is also
 shown (dashed line) as explained in the text.} \label{prospects}
\end{figure}

Data collection is currently in progress and some strategies are
being considered to further reduce the low energy background.
Another 50 \% reduction from 4 keV to 10 keV (which could be
reasonably expected) would allow to explore practically all the
DAMA region in 1~kg~y of exposure. In the case of reducing the
background down to the flat level of 0.04 c/kg/keV/day (currently
achieved by IGEX for energies beyond 20 keV), that region would be
widely surpassed. In Figure~\ref{prospects} we plot the exclusions
obtained with a flat background of 0.1 c/kg/keV/day (dot-dashed
line) and of 0.04 c/kg/keV/day (solid line) down to the current 4
keV threshold, for an exposure of 1~kg~year. As can be seen, the
complete DAMA region (m=$52^{+10}_{-8}$ GeV,
$\sigma^p$=($7.2^{+0.4}_{-0.9}$)x10$^{-9}$ nb) could be tested
with a moderate improvement of the IGEX performances.

A new experimental project on WIMP detection using larger masses
of Germanium of natural isotopic abundance (GEDEON, GErmanium
DEtectors in ONe cryostat) is planned. It will use the technology
developed for the IGEX experiment and it would consist of a set of
$\sim$1 kg germanium crystals, of a total mass of about 28 kg,
placed together in a compact structure inside one only cryostat.
This approach could benefit from anticoincidences between crystals
and a lower components/detector mass ratio to further reduce the
background with respect to IGEX. A detailed study is in progress
to assess the physics potential of this device, but it can be
anticipated that a flat background of 0.002 c/kg/keV/day down to a
threshold below 4 keV is a reasonable estimate. The exclusion plot
which could be expected with such proviso for 24 kg y of exposure
is shown in the Figure~\ref{prospects}. Moreover, following the
calculations presented in \cite{Cebrian:2001}, GEDEON would be
massive enough to search for the WIMP annual modulation effect and
explore positively an important part of the WIMP parameter space
including the DAMA region.

\section*{Acknowledgements}
The Canfranc Astroparticle Underground Laboratory is operated by
the University of Zaragoza under contract No. AEN99-1033. This
research was funded by the Spanish Commission for Science and
Technology (CICYT), the U.S. National Science Foundation, and the
U.S. Department of Energy. The isotopically enriched $^{76}$Ge was
supplied by the Institute for Nuclear Research (INR), Moscow, and
the Institute for Theoretical and Experimental Physics (ITEP),
Moscow.

\newpage

\newpage

\newpage

\newpage

\newpage

\begin{table}[htb]
\begin{center}
\begin{tabular}[h]{cccccc}
\hline \multicolumn{1}{r}{{\bf E (keV)}}
                 & \multicolumn{1}{r}{{\bf counts}}

                 & \multicolumn{1}{r}{{\bf E (keV)}}
                 & \multicolumn{1}{r}{{\bf counts}}

                 & \multicolumn{1}{r}{{\bf E (keV)}}
                 & \multicolumn{1}{r}{{\bf counts}}
                  \\
\hline
\small
4.5 &   18  &   19.5    &   4   &   34.5    &   4   \\ 5.5 &   25
&   20.5    &   5   &   35.5    &   4   \\ 6.5 &   16  &   21.5 &
1   &   36.5    &   6   \\ 7.5 &   11  &   22.5    &   4   & 37.5
&   3   \\ 8.5 &   23  &   23.5    &   4   &   38.5    & 3   \\
9.5 &   9   &   24.5    &   4   &   39.5    &   3   \\ 10.5 &   12
&   25.5    &   4   &   40.5    &   5   \\ 11.5    &   17 &   26.5
&   4   &   41.5    &   4   \\ 12.5    &   12  & 27.5    &   9   &
42.5    &   0   \\ 13.5    &   7   &   28.5 &   4   &   43.5    &
2   \\ 14.5    &   6   &   29.5    &   3 &   44.5    &   3   \\
15.5    &   6   &   30.5    &   2   & 45.5    &   5   \\ 16.5    &
8   &   31.5    &   2   &   46.5 &   2   \\ 17.5    &   6   &
32.5    &   1   &   47.5    &   3
\\ 18.5    &   1   &   33.5    &   1   &   48.5    &   4   \\

\hline
\end{tabular}
\caption{Low-energy data from the IGEX RG-II detector (Mt $=$
80~kg~d).} \label{tab-ig-1}
\end{center}
\end{table}

\newpage

\newpage

\newpage

\newpage

\end{document}